\documentclass{article}

\usepackage{spconf,amsmath,amssymb,graphicx,epsfig, cite,algorithm,algorithmic,epstopdf, url}
\usepackage[utf8]{inputenc}
\usepackage[mathscr]{euscript}
\usepackage{color}
\usepackage{bm}
\usepackage[keeplastbox]{flushend}
\usepackage[title]{appendix}
\usepackage{relsize}

\def\minwrt[#1]{\underset{#1}{\text{minimize }}}
\def\argminwrt[#1]{\underset{#1}{\text{arg min }}}
\def\maxwrt[#1]{\underset{#1}{\text{maximize }}}
\def\maxemphwrt[#1]{\underset{#1}{\text{\emph{maximize} }}}
\newcommand{\ett}{{\bf 1}}

\newcommand{\diag}{{\rm diag}}

\newtheorem{proposition}{Proposition}
\newtheorem{proof}{Proof}

\newcommand{\norm}[1]{\left\lVert#1\right\rVert}

\newcommand{\trace}[1]{\text{tr}\left(#1\right)}

\def\ccM{{\mathcal{M}}}

\def\ccS{{\mathcal{S}}}

\def\RC{{\mathbb{C}}}

\def\RR{{\mathbb{R}}}

\newcommand{\Omtspec}{T}
\newcommand{\costfunc}{c}

\newcommand{\odotproduct}[2]{\underset{#1}{\overset{#2}{{\bigodot}}} \ }

        \makeatletter
        \def\fps@eqnfloat{!t}
        \def\ftype@eqnfloat{4}
        
        \newenvironment{eqnfloat*}
               {\@dblfloat{eqnfloat}}
               {\end@dblfloat}
        \makeatother
\renewenvironment{thebibliography}[1]{%
 \begin{oldthebibliography}{#1}%
 \setlength{\parskip}{0ex}%
 \setlength{\itemsep}{0ex}%
}%
{%
\end{oldthebibliography}%
}%

\begin{document}
%
\title{Non-coherent Sensor Fusion via \\entropy regularized 
optimal mass transport}


%
\name{Filip Elvander$^*$, Isabel Haasler$^\dagger$, Andreas Jakobsson$^*$, and Johan Karlsson$^\dagger$\thanks{This work was supported in part by the Swedish Research Council, Carl Trygger's foundation, and the Royal Physiographic Society in Lund.}}

\address{
$^*$Div. of Mathematical Statistics, Lund University, Sweden\\
$^\dagger$Dept. of Mathematics, KTH Royal Institute of Technology, Sweden\\
emails: \{filipelv, aj\}@maths.lth.se,  \{haasler, johan.karlsson\}@math.kth.se\vspace{-4pt}}



\maketitle


\begin{abstract}
This work presents a method for information fusion in source localization applications. 
The method utilizes the concept of optimal mass transport in order to construct estimates of the spatial spectrum using a convex barycenter formulation. We introduce an entropy regularization term to the convex objective, which allows for low-complexity iterations of the solution algorithm and thus makes the proposed method applicable also to higher-dimensional problems. 
We illustrate the proposed method's inherent robustness to misalignment and miscalibration of the sensor arrays using numerical examples of localization in two dimensions.
\end{abstract}
\vspace{2mm}
\begin{keywords}
Optimal mass transport, Entropy regularization, Target localization, Sensor fusion, Non-coherent processing.
\end{keywords}
\section{Introduction}
Source localization using an array of sensors is a well established  and still highly active field of research in signal processing, with applications in areas such as radar, sonar, and audio processing \cite{KrimV96, ChenYH02_19, StoicaM05, DeyA18}. 
The localization problem is 
often formulated as a spatial spectral estimation problem, where signal sources are identified with peaks of power in the spectral representation. In contrast to the closely related problem of direction of arrival (DoA) estimation, where one aims to find the directions of incoming planar waves from far field sources, the problem of localization also includes the estimation of the range (i.e., distance to the sources) and thus needs to utilize that the incoming wavefronts are non-planar. 
Classical methods, originally used mainly for DoA estimation, such as beamforming techniques \cite{Capon69} and subspace methods like MUSIC \cite{Schmidt79} and ESPRIT \cite{RoyK89_37}, may be readily extended to the localization problem, with more recent efforts including works exploiting spectral sparsity \cite{StoicaBL11_59b, AdalbjornssonKBAJ16_24, YangX16_64}.
Many of these methods rely on the assumption of well-calibrated arrays, i.e., known sensor positions, in order to produce reliable estimates. However, this assumption may in some applications be violated due to, e.g., calibration errors, perturbations of the sensor localizations, or synchronization errors. Therefore, considerable effort has been directed towards formulating estimators robust to such calibration errors (see, e.g., \cite{YangS95_43, SomasundaramJ14, RubsamenP13_61, LorenzB05_53, PesaventoGW02_50}).
In some of the approaches to address this problems, techniques have been developed that divide the available sensors into subarrays, each assumed to be perfectly calibrated, although relative errors between the different subarrays are allowed for \cite{SeeG04_52}. 
The DoA estimation and/or localization is then performed by minimizing measures of discrepancy between the measurements, i.e., the estimated joint covariance matrix of the complete set of sensors, and the assumed array geometry \cite{PesaventoGW02_50}.

In this work, we instead consider localization by fusing information from separate sensor arrays, where we only assume to have access to measurements in the form of covariance matrices for the individual arrays, so-called non-coherent processing (see, e.g., \cite{WaxK85_33, RiekenF04_54, SuleimanPPZ18_66, KimHE15_22}), i.e., we assume no information about cross-correlation between different arrays. Such a scenario may arise in distributed sensor networks, wherein sensor nodes transmit information to a central processing unit  that performs the estimation \cite{Kaplan06_42,LiWHS02_19}. In such settings, it is common to have restrictions on the communication bandwidth, which requires transmitting only aggregated data \cite{Kaplan06_42}, e.g., in the form of covariance matrices. Also, as the separate sensor arrays may not be jointly synchronized, cross-correlations cannot be estimated from the raw data. 

Extending on the work presented in \cite{ElvanderHJK18_eusipco}, which proposes a robust method for DoA sensor fusion using optimal mass transport (OMT), we here generalize this approach to the  source localization problem.
One of the main ideas is  to use OMT as a notion of distance between spectra (see, e.g., \cite{GeorgiouKT09_57, ElvanderJK18_66, ElvanderJK18_icassp}), and in particular, the proposed method performs sensor fusion by computing the corresponding barycenter solutions based on incomplete measurements.
In addition, to make the scheme computationally efficient, we introduce an entropy regularization term in the problem formulation and show how to solve the resulting optimization problem iteratively, similar to the popular Sinkhorn iteration scheme \cite{Cuturi13} (see also \cite{KarlssonR17_10}). The robustness of the proposed method is demonstrated in numerical examples.

%
\section{Background}\label{sec:background}
\subsection{Near field localization} \label{ssec:DOA}
Consider an array consisting of $p$ sensors located at $y_k \in \RR^d$, for $k = 1,2,\ldots, p$, and assume that point-like sources, located in a space $\ccS \subset \RR^d$, emit spherical waves impinging on the sensors. Then, the spatial distribution of the energy of the source signals may be representated by a non-negative function, or measure, $\Phi$ on $\mathcal{S}$, referred to as the power spectrum. We will denote the space of such functions by $\ccM_+(\ccS)$. Letting $\lambda$ denote the wavelength of the impinging waves, the covariance matrix of the sensor measurements is $R = \Gamma(\Phi)$, where the linear operator $\Gamma: \ccM(\ccS) \to \RC^{p\times p}$ is given by (see, e.g., \cite{Georgiou05_50})
\begin{align}\label{eq:def_gamma}
\Gamma(\Phi)\triangleq\int_{\ccS} a(x)\Phi(x)a(x)^H dm(x),
\end{align}
where $a(x) \in \RC^p$ is the array manifold vector \cite{JohnsonD93}, i.e.,
\begin{equation}
a(x)=\begin{pmatrix}\frac{1}{\norm{y_k-x}_2^{(d-1)/2} }e^{-{2\pi
i}\!\frac{\norm{y_k-x}_2}{\lambda}} \end{pmatrix}_{k=1}^p,
\label{eq:array_mainifold_vector}
\end{equation}
with $dm(x)$ denoting the Lebesgue measure. In other words, a positive semi-definite matrix $R$ is a valid covariance for a sensor array if and only if $R = \Gamma(\Phi)$, for some $\Phi \in \ccM_+(\ccS)$. From this, it may be noted that performing source localization based on sensor measurements corresponds to inferring a power spectrum $\Phi \in \ccM_+(\ccS)$ from observations of the array covariance matrix $R$. 

We will here consider a scenario wherein several sensor arrays measure the superposition of the impinging waves, i.e., when covariance matrices $R_j$, for $j = 1,\ldots,J$, are available, corresponding to $J$ separate sensor arrays. For the case when all arrays are calibrated, i.e., when the position of the sensors are known perfectly, source localization thus corresponds to finding a single power spectrum $\Phi$ that is consistent with the observed matrices, i.e., $\Gamma_j(\Phi) = R_j$, for $j = 1,\ldots,J$, where $\Gamma_j$ is the linear operator corresponding to array $j$. However, for the case of miscalibration, the (assumed) operators $\Gamma_j$ contain errors, which may result in there being no single $\Phi \in \ccM_+(\ccS)$ consistent with all covariance matrices. We will here adress this problem by utilizing OMT in order to induce robustness to the sensor fusion problem.
\subsection{Optimal mass transport}
Let $\Phi_0$ and $\Phi_1$ be two non-negative mass distributions on the space ${\ccS}$. A transport plan $M \in \ccM_+(\ccS^2)$ for such a pair is a non-negative distribution on $\ccS^2$, with $\ccS^2 = \ccS\times \ccS$, where $M(x_0, x_1)$ describes the amount of mass transported from location $x_0\in \ccS$ to location  $x_1\in \ccS$.
By defining a cost function $\costfunc:\ccS^2 \to \RR$ describing the cost of moving mass, we may define the total cost associated with a transport plan as
\begin{align}
	\Psi(M) \triangleq \int_{\ccS^2} \costfunc(x_0,x_1)M(x_0,x_1)dm(x_0)dm(x_1).
\end{align}
The OMT problem is then to find the feasible transport plan from $\Phi_0$ to $\Phi_1$ incurring minimal transportation cost (see, e.g., \cite{Villani08}), i.e., to solve
\begin{align} 
\Omtspec(\Phi_0,\Phi_1)\triangleq\!\!\min_{M\in \ccM_+(\ccS^2)}\;& \Psi(M) \label{eq:omt}\\
\mbox{ subject to }\; &\Phi_0(x_0)=\int_\ccS M(x_0,x_1)dm(x_1)\nonumber\\
&\Phi_1(x_1)=\int_\ccS M(x_0,x_1)dm(x_0), \nonumber 
\end{align}
where the constraints ensure that the transport plan is valid, i.e., matches the marginals $\Phi_0$ and $\Phi_1$.
The suitability of utilizing OMT costs as measures of distance has previously been considered for imposing metric structure on the space of power spectra \cite{GeorgiouKT09_57}, as well as to induce distances and smooth interpolants for covariance matrices \cite{ElvanderJK18_66}.
In this work, the optimal transport cost $\Omtspec(\Phi_0,\Phi_1)$ will be used for interpolating between mass distributions $\Phi_0$ and $\Phi_1$. It is worth noting that the OMT cost quantifies location errors based on the distances on the underlying space, in this case $\ccS \subset \RR^d$, thereby allowing for offsets due to, e.g., calibration errors. This is in contrast to standard metrics (e.g., $L_1$, $L_2$) which only compare functions point by point without taking the geometry of the underlying space, i.e., relative location of the points, into account.
\section{Proposed method}%
Consider the localization scenario with the source signals impinging on a set of $J$ arrays, with array manifold vectors $a_j(x)$, generating the covariance matrices $R_j$, $j=1,\dots,J$. Given the array geometries and covariance matrix estimates of each individual array, we aim to estimate the spatial spectrum $\Phi$. 
That is, we assume that each sensor array is calibrated within itself, but there might be global calibration errors resulting in 
inaccurate knowledge of each sensor array's position and orientation.
For the far field case, i.e., DoA estimation, an OMT formulation for fusing information from different sensor arrays has been proposed in \cite{ElvanderHJK18_eusipco}. There, the estimate was formed as the spectrum closest, in the OMT sense, to a set of $J$ spectra, each consistent with a corresponding covariance matrix.
Correspondingly, in the localization setting the spatial spectrum $\Phi$ may be found as the solution to the convex barycenter problem
\begin{equation}\label{eq:sensor_fusion}
\begin{aligned}
\minwrt[\Phi \in \ccM_+(\ccS),\Phi_j \in \ccM_+(\ccS)] &\quad \sum_{j = 1}^J \Omtspec(\Phi,\Phi_j) \\
\text{subject to } \qquad &\quad \Gamma_j(\Phi_j) = R_j,\; j = 1,2,\ldots,J,
\end{aligned}
\end{equation}
where $\Gamma_j$ denotes the linear operator \eqref{eq:def_gamma} of array $j$.
In order to find practical solutions to the barycenter problem in \eqref{eq:sensor_fusion}, the space $\ccS$ may be discretized as to be represented by $n$ points $x_k \in \ccS$. Accordingly, the marginal spatial spectra are represented by the vectors $\Phi_j\in \RR_+^n$, and the cost function and transport plans are represented by matrices $C = \left[ \costfunc_{k,\ell}\right] \in \RR^{n\times n} $ and $M = \left[ m_{k,\ell}\right]\in \RR_+^{n\times n}$, respectively, where $\costfunc_{k,\ell} = \costfunc(x_k,x_\ell)$, and $m_{k,\ell}$ denotes the mass transported from $x_k$ to $x_\ell$.
Although such a discretization yields finite-dimensional linear programs, finding an optimal transport plan can be computationally demanding due to the large number of variables.

For discretizations of the standard OMT problem \eqref{eq:omt}, this was in \cite{Cuturi13} addressed by regularizing the problem by introducing an entropy term to the objective function, yielding the strictly convex program
\begin{equation}
\begin{aligned} 
T_{\epsilon}(\Phi_0,\Phi_1) = \quad \minwrt[M \in \RR_+^{n\times n}]  \quad & \trace{C^T M} + \epsilon D(M) \label{eq:omt_regularized} \\
\mbox{ subject to } \quad &\Phi_0=M\ett \:,\; \Phi_1=M^T \ett ,
\end{aligned}
\end{equation}
where $D(M) = \sum_{k,\ell=1}^n \left( m_{k,\ell} \log(m_{k,\ell}) - m_{k,\ell} + 1 \right)$ is an entropy term, $\epsilon>0$ a small constant, and  $\ett \in \RR^n$ denotes a vector of ones. In \cite{Cuturi13}, it was shown that this problem may be solved in a computationally efficient way, using so-called Sinkhorn iterations.
\subsection{Efficient sensor fusion}
In order to arrive at a computationally efficient scheme for computing spectral barycenters, we propose to regularize the problem in \eqref{eq:sensor_fusion} according to
\begin{equation}
\begin{aligned}
\minwrt[\Phi, \Phi_j, \Delta_j] \;\ & \sum_{j=1}^J  \left(  T_\epsilon(\Phi, \Phi_j) + \gamma \|\Delta_j\|_2^2 \right) \\
\textrm{ subject to } \;\  &  A_j \Phi_j  = \Delta_j + r_j \quad \mbox{for } j=1,\dots,J.
\end{aligned} \label{eq:barycenter_discrete}
\end{equation}
Here, $r_j = \text{vec}(R_j) \in \RC^{p_j^2}$, for $j = 1,\ldots,J$, are the vectorizations of the covariances matrices, and $A_j \in \RC^{p_j^2 \times n}$ are matrix representations of the discretized operators $\Gamma_j$. Also, the vectors $\Delta_j \in \RC^{p_j^2}$ allow for noisy covariance matrix estimates, penalized by the standard $\ell_2$-norm, as controlled by the positive penalty parameter $\gamma$.
The addition of the entropy term, $D(M)$, renders the problem strictly convex and allows for efficiently computing the barycenter $\Phi$, using the observation in \cite{KarlssonR17_10}, that Sinkhorn iterations are equivalent to a block coordinate ascent in a Lagrange dual of the entropy regularized transport problem.
This then allows for increasing the dimensionality of the sensor fusion problem, i.e., making it feasible to include also range estimation or considering higher spatial dimensions. The extended Sinkhorn iterations, constituting a globally convergent solution algorithm for \eqref{eq:barycenter_discrete}, are detailed in the following proposition.
%
%
%
\begin{algorithm}[t]\caption{Sinkhorn-Newton method for the optimal mass transport barycenter problem with partial information.}\label{alg:barycenter_sinkhornnewton}
	\begin{algorithmic}
		\STATE Require: Initial guess $\lambda_j, v_j,$  and let $u_j = \exp(A_j\lambda_j/\epsilon)$ for $j=1,\ldots,J$
		\WHILE{Sinkhorn not converged}
		\FOR{j=1:J}
		\WHILE{Newton not converged}
		\STATE $f \leftarrow A_j \left(u_j \odot Kv_j \right) - r_j + 1/(2\gamma) \lambda_j$
		\STATE $df \leftarrow (1/\epsilon) A_j \diag( u_j \odot Kv_j ) A_j^T + 1/(2\gamma)I $
		\STATE $\Delta\lambda \leftarrow - df\backslash f$
		\STATE $\lambda_j \leftarrow \lambda_j+\sigma\Delta\lambda$, with $\sigma$ determined by a linesearch
		\STATE $u_j \leftarrow \exp(A_j\lambda_j/\epsilon)$
		\ENDWHILE
		\ENDFOR
		\STATE $\Phi \leftarrow \left( \odotproduct{j=1}{J}  (K^T u_j) \right)^{1/J}$
		\FOR{j=1:J}
		\STATE $v_j \leftarrow \Phi./ (K^T u_j) $
		\ENDFOR
		\ENDWHILE
		\RETURN $\Phi$ 
	\end{algorithmic}
\end{algorithm}
%
%
\begin{proposition}
	A block coordinate ascent scheme for the dual of the barycenter problem \eqref{eq:barycenter_discrete} is to iteratively perform
	\begin{enumerate}
		\item For $j=1,\ldots,J$, let $\lambda_j$ be the solution to
		\begin{equation}
		A_j \left(u_j(\lambda_j) \odot Kv_j \right) - r_j + \frac{1}{2\gamma} \lambda_j = 0, \label{eq:barycenter_sinkhorn_maximization}
		\end{equation}
		where
		\begin{equation*}
		u_j(\lambda_j) = \exp(A_j^T \lambda_j / \epsilon).
		\end{equation*}
		\item Update 
		\begin{equation*}
		v_j= y./ (K^T u_j(\lambda_j))  \quad \text{for } j=1,\dots,J, 
		\end{equation*}
		where
		\begin{equation*}
		y = \left( \odotproduct{j=1}{J}  (K^T u_j(\lambda_j)) \right)^{1/J}.
		\end{equation*}
	\end{enumerate}
	Here, $K = \exp(-C/\epsilon)$, and $\exp(\cdot)$,  $./$, and $\bigodot$ denotes the elementwise exponential function, elementwise division, and elementwise product, respectively.
	The barycenter solution is then given by $\Phi=y$, where $y$ is the limit point of the iterations.
	Note that we here convert all complex quantities to an equivalent real-valued representation such that the vectorized covariances, $r_j$, are mapped according to
\begin{align*}
	r_j \mapsto \left[\begin{array}{cc} \mathfrak{Re}\left( r_j  \right) & \mathfrak{Im}\left( r_j  \right) \end{array}\right]^T,
\end{align*}
where $\mathfrak{Re}(\cdot)$ and $\mathfrak{Im}(\cdot)$ denote the real and imaginary parts, respectively, and where $\Delta_j$ and the discretized operators $A_j$ are converted as to be consistent with this.
	 \label{thm:barycenter_sinkhorn}
\end{proposition}
A proof of the proposition may be found in Appendix \ref{appendix:proof}.
It may be noted that for the noise-free case with complete observations of the underlying spectra, i.e., $\Delta_j = 0$ and $A_j = I$, as well as $J = 2$, one obtains the standard Sinkhorn iterations. 
The roots of \eqref{eq:barycenter_sinkhorn_maximization} are determined by the means of a Newton method, using the Jacobian
\begin{equation*}
\frac{1}{\epsilon} A_j \diag( u_j(\lambda_j) \odot Kv_j) A_j^T + \frac{1}{2\gamma}I.
\end{equation*}
The complete estimation method for computing the spectral barycenter is summarized in Algorithm \ref{alg:barycenter_sinkhornnewton}.
%
%
\begin{figure}[t]
        \centering
        \vspace{1mm}
            \includegraphics[width=0.45\textwidth]{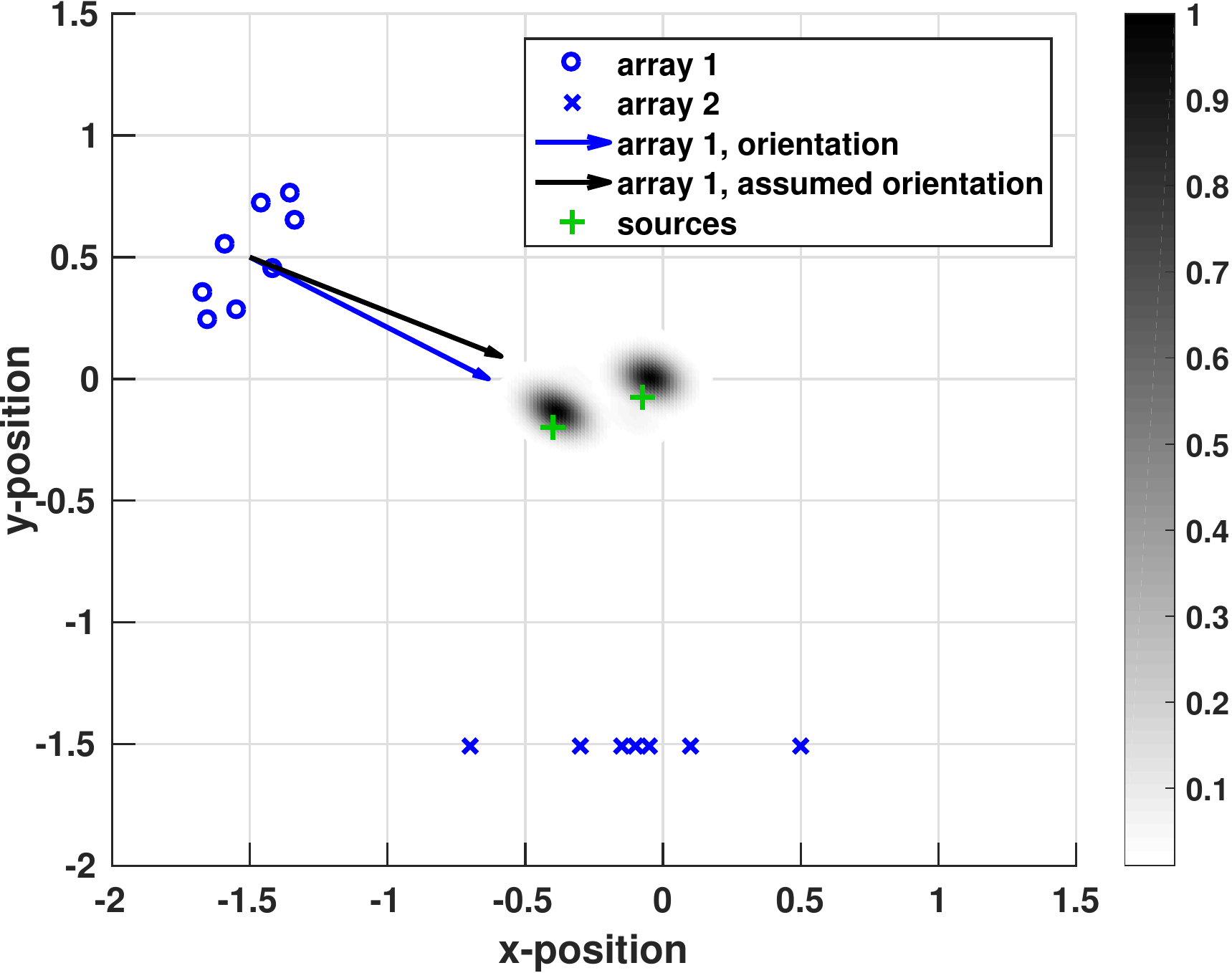}
            \vspace{.75mm}
           \caption{Spectral estimate as given by the proposed estimator in \eqref{eq:barycenter_discrete}. The alignment error is 6.7 degrees.}
            \label{fig:sensor_fusion_2D_OMT}
\end{figure}
%
%
\begin{figure}[t]
        \centering
        \vspace{1mm}
            \includegraphics[width=0.45\textwidth]{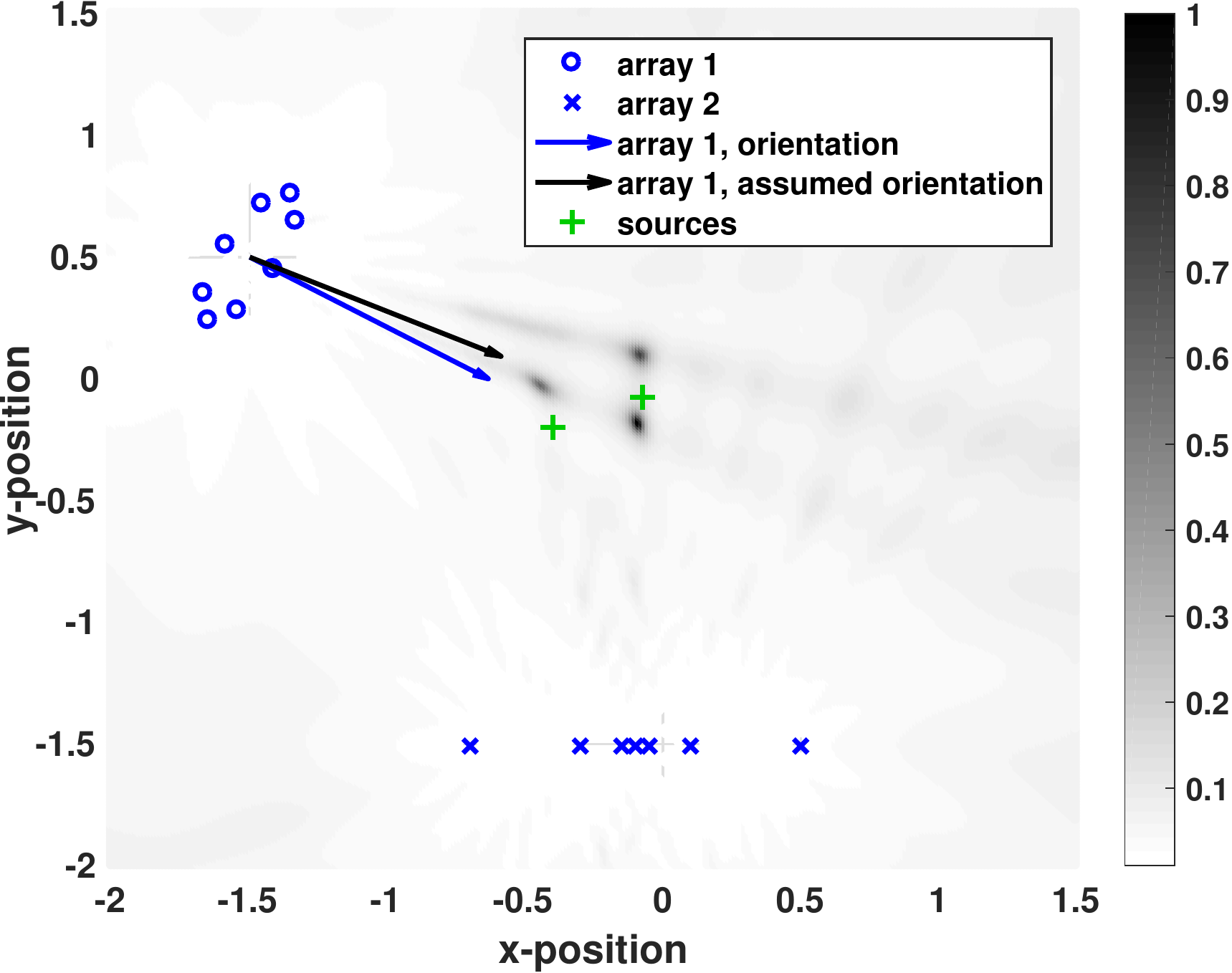}
            \vspace{.75mm}
           \caption{Pseudo-spectrum as given by non-coherent MUSIC. The alignment error is 6.7 degrees.}
            \label{fig:sensor_fusion_2D_music}
\end{figure}
%
\begin{figure}[t!]
        \centering
        \vspace{1mm}
            \includegraphics[width=0.45\textwidth]{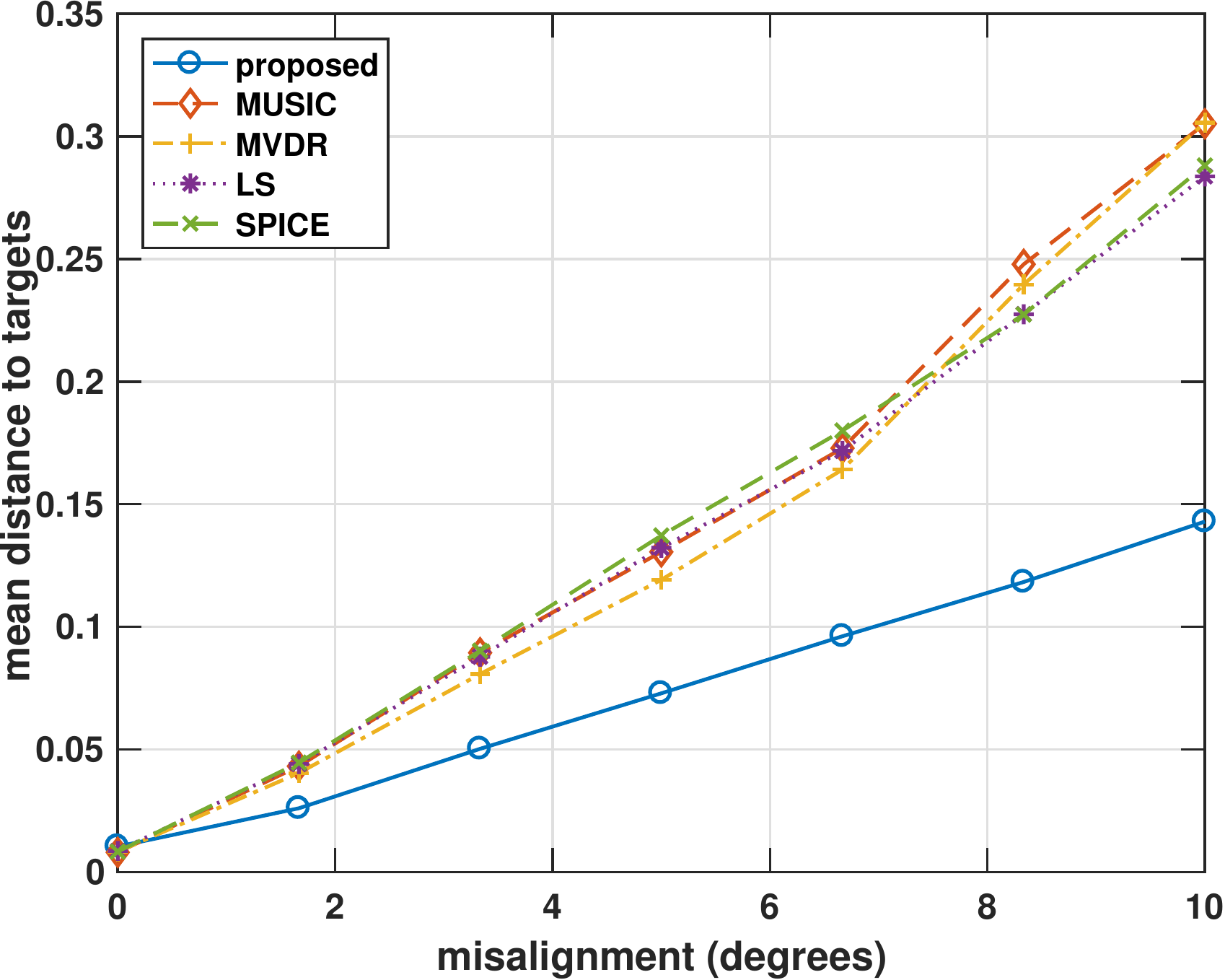}
            \vspace{.75mm}
           \caption{Average distance from estimates to signal sources, as function of the misalignment angle.}
            \label{fig:sensor_fusion_2D_simulation}
\end{figure}
%
\section{Numerical results}
In this section, we illustrate the proposed method's behavior in sensor fusion scenarios for localization in 2D, demonstrating the robustness to array alignment errors that the OMT criterium induces. Specifically, we consider two uncorrelated sources impinging on two asynchronous sensor arrays; one ellipsoidal shaped array consisting of 8 sensors, and one linear array consisting of 7 sensors. To simulate misalignment between the two arrays, we introduce an unknown rotation to the ellipsoidal array. The scenario, as shown in Figure~\ref{fig:sensor_fusion_2D_OMT}, is then used in a simulation study. Varying the misalignment, i.e., the rotation angle, between 0 and 10 degrees, we for each considered angle perform 100 Monte Carlo simulations, were the locations of the signal sources are randomized uniformly on the square $[-0.5,0.5]\times[-0.5,0.5]$.
We let the source signals be uncorrelated circularly symmetric Gaussian white noises and add an uncorrelated Gaussian noise to the sensors. The source signals have variance 100 and the variance of the sensor noise is 1. The wavelength of the impinging waves is twice that of the smallest sensor spacing in the linear array.
Estimating the covariance matrices of the separate sensor arrays using the sample covariance matrix and 500 signal snapshots, we apply the proposed estimator in \eqref{eq:barycenter_discrete} to the covariance matrix estimates and record the distance from the location estimates to the actual target positions. Throughout, we use the parameters $\gamma = 0.01$ and $\epsilon = 0.005$, and the cost function $\costfunc(x_k,x_\ell) = \norm{x_k-x_\ell}_2^2$, for grid points $x_k,x_\ell \in \RR^2$.

As comparison, we also obtain estimates using the non-coherent MUSIC and MVDR estimators, as described in \cite{RiekenF04_54}, as well as the least-squares (LS) estimator from \cite{LuoYWH17_270}, and the non-coherent SPICE estimator from \cite{SuleimanPPZ18_66}. Figures~\ref{fig:sensor_fusion_2D_OMT} and \ref{fig:sensor_fusion_2D_music} show the estimates obtained in one simulation using the proposed and MUSIC estimators, respectively. The misalignment angle is 6.7 degrees. As can be seen, for the proposed estimator, the misalignment only causes a small perturbation on the estimated source locations, demonstrating the ability of the OMT formulation to perform smooth interpolation on the underlying domain. In contrast, the MUSIC (pseudo-) spectral estimate contains spurious peaks, as well as larger deviations from the true source locations. The robustness of the proposed estimator is further demonstrated in Figure~\ref{fig:sensor_fusion_2D_simulation}, displaying the results from the simulations. From this, it may be noted that the proposed estimator deviates slightly more from the ground truth for the non-perturbed case as compared to the other methods. However, the robustness to misalignment is significantly higher; the rate of error increase is considerably lower than for the other methods.
\begin{appendices}
\section{Proof of Proposition}\label{appendix:proof}
\begin{proof}
In the following, we denote  the set $\{M_1,\ldots,M_J\}$ by $\mathfrak{M}$, and similarly for $\Delta$, $\lambda$ and $\mu$. The Lagrangian of \eqref{eq:barycenter_discrete} with the dual variables $\lambda$ and $\mu$ is
\begin{align}
	L(\Phi, & \mathfrak{M} ,\Delta,\lambda,\mu) =\sum_{j=1}^J \bigg(   \trace{C^T M_j } + \epsilon D(M_j) +  \gamma \|\Delta_j\|^2  \nonumber \\
   & + \lambda_j^T (\Delta_j + r_j - A_j M_j^T \mathbf{1} ) + \mu_j^T (\Phi - M_j \mathbf{1} ) \bigg).\label{eq:barycenter_lagrangian}
\end{align}
Minimizing the Lagrangian with respect to $M_j$ gives the expression for the mass transport matrix solutions,
\begin{equation}
M_j = \diag(u_j ) K \diag( v_j), \label{eq:barycenter_M_j}
\end{equation} 
with $K=\exp(-C/\epsilon)$, $u_j=\exp(A_j^T\lambda_j/\epsilon)$ and $v_j=\exp(\mu_j/\epsilon)$, for $j=1,\ldots,J$.
Further, minimizing \eqref{eq:barycenter_lagrangian} with respect to the variables $\Phi$ and $\Delta$, one gets a dual problem to \eqref{eq:barycenter_discrete}, formulated as
\begin{equation}
\begin{aligned}
\maxwrt[\lambda_j,\mu_j] & \sum_{j=1}^J L(\Phi^*,\mathfrak{M}^*,\Delta^*,\lambda,\mu)_j \label{eq:barycenter_dual}\\
\mbox{subject to } & \sum_{j=1}^J \mu_j = 0,
\end{aligned}
\end{equation}
where
\begin{equation*}
\begin{aligned}[t] L(\Phi^*,{\mathfrak{M}}^*,\Delta^*,\lambda,\mu)_j = ( \lambda_j^T r_j - \epsilon e^{(A_j^T\lambda_j)^T/\epsilon} e^{-C/\epsilon } e^{ \mu_j /\epsilon}  \\ 
- \frac{1}{4 \gamma} \lambda_j^T \lambda_j + \epsilon n^2 ). \end{aligned}
\end{equation*}
A block coordinate ascent method is to maximize \eqref{eq:barycenter_dual} iteratively with respect to $\lambda$ and $\mu$. Maximization with respect to $\lambda_j$ requires
\begin{equation*}
r_j - A_j \diag( e^{-C/\epsilon} e^{\mu_j/\epsilon} ) e^{A_j^T\lambda_j /\epsilon} - \frac{1}{2\gamma} \lambda_j = 0,
\end{equation*}
which gives the first step \eqref{eq:barycenter_sinkhorn_maximization}.
To maximize \eqref{eq:barycenter_dual} with respect to $\mu_j$, consider the corresponding Lagrangian with dual variable $y$, given by
\begin{equation*}
L(\mu,y) = - \epsilon \sum_{j=1}^J   e^{(A^T\lambda_j)^T/\epsilon} e^{-C/\epsilon} e^{\mu_j /\epsilon}  + y \sum_{j=1}^J \mu_j .
\end{equation*}
Maximization with respect to $\mu_j$ requires
\begin{equation*}
- \diag( e^{-C^T/\epsilon} e^{A_j^T\lambda_j/\epsilon} )  e^{\mu_j /\epsilon} + y = 0 \
 \Leftrightarrow \ v_j = y./ (K^T u_j).
\end{equation*}
%
Further, due to the constraint in \eqref{eq:barycenter_dual}, it follows
\begin{equation*}
\ett = \odotproduct{j=1}{J} v_j = \odotproduct{j=1}{J} y./ (K^T u_j) \	\Leftrightarrow \ y = \left( \odotproduct{j=1}{J}(K^T u_j) \right)^{1/J}.
\end{equation*}
This completes the second step of the method.
Finally, with \eqref{eq:barycenter_M_j}, we get the expression for the barycenter marginal
\begin{equation*}
\Phi = M_j^T \mathbf{1} = v_j \odot (K^T u_j) = y. 
\end{equation*}
\end{proof}
\end{appendices}
\bibliographystyle{IEEEbib}
\bibliography{IEEEabrv,ICASSP_2019_arXiv}

\end{document}